\documentclass{article}
\usepackage{eufrak}
\usepackage[mathscr]{eucal}

\def\be{\begin{equation}}
\def\ee{\end{equation}}
\def\bea{\begin{eqnarray}}
\def\eea{\end{eqnarray}}
\def\beqa{\begin{eqnarray}}
\def\eeqa{\end{eqnarray}}
\def\beq{\begin{equation}}
\def\eeq{\end{equation}}
\def\beqal{\begin{eqnarray}\label}
\def\beql{\begin{equation}\label}

\def\R{\mbox{\rm I\kern-.18em R}}
\def\Rq{\R^4}
\def\P{\mbox{\rm I\kern-.18em P}}
\def\uno{\mbox{1 \kern-.59em {\rm l}}}
\def\Ds{\ {\big / \kern-.70em D}}
\def\ds{\big / \kern-.90em {\ \p} }

\def\cDs{\ {\big / \kern-.70em {\cal D}}}
\def\Z{{Z \kern-.45em Z}}
\def\Q{{\kern .1em {\raise .47ex \hbox{$\scriptscriptstyle |$}}
\kern -.35em {\rm Q}}}

\def\Tr{\mbox{\rm Tr}}

\def\p{\partial}

\def\m{\mu}
\def\n{\nu}

\def\L{\Lambda}

\def\ie{{\it i.e.}}

\def\1{\dot{1}}
\def\2{\dot{2}}

\def\ie{{\it i.e. }}

\def\Journal#1#2#3#4{{#1} {\bf #2}, #3 (#4)}

\def\NPB{{\em Nucl. Phys.} B}
\def\PLB{{\em Phys. Lett.}  B}
\def\PRL{\em Phys. Rev. Lett.}
\def\PRD{{\em Phys. Rev.} D}

\font\mybb=msbm10 at 12pt

\def\bb#1{\hbox{\mybb#1}}
\def\Z {\bb{Z}}
\def\C {\bb{C}}

\begin{document}

\title{\large \sc  Multi--Instantons, Supersymmetry  and
 Topological Field Theories}
\author{
{\sc Diego Bellisai}\\
{\sl Dipartimento di Fisica ``G. Galilei'', Universit\`a di Padova}\\
{\sl and I.N.F.N. Sezione di Padova, Via Marzolo 8, 35131 Padova, Italy}\\
\and
{\sc Francesco Fucito, Alessandro Tanzini,  Gabriele Travaglini}\\
{\sl I.N.F.N. Sezione di Roma II,}\\   
{\sl Via della Ricerca Scientifica, 00133 Roma, Italy}
}
\maketitle
\begin{center}
{\large \bf ABSTRACT}\\
\end{center}
In this letter we 
argue that instanton--dominated Green's functions in $N=2$ Super Yang--Mills
theories can be equivalently computed either using the so--called constrained 
instanton method or making reference to the topological twisted version of the 
theory. Defining an appropriate BRST operator
(as a supersymmetry plus  a gauge variation), we also show that 
the expansion coefficients of the Seiberg--Witten effective action 
for the low--energy degrees of freedom can be written as integrals 
of total derivatives over the moduli space of self--dual gauge connections.

\vskip 1.0cm
\setcounter{section}{0}
\section{Introduction}
Much progress has been made in recent years in understanding 
non--perturbative phenomena in globally supersymmetric (SUSY) gauge theories.
In a seminal paper \cite{sw}, Seiberg and Witten  calculated all the 
non--perturbative contributions to the holomorphic 
effective action for a Super Yang--Mills (SYM) theory
with $N=2$ global supersymmetry 
using certain ans\"atze dictated by physical intuitions.
To check the correctness of this result, the 
contributions to the  prepotential ${\cal F}$
coming from instantons of winding number one and two
were computed directly  
in the Coulomb phase of the  theory, 
by using a saddle point approximation for the calculation of 
the relevant correlators \cite{fp,DKM,ft}. 
The results were found to be in agreement with the  expression  
of ${\cal F}$ derived in \cite{sw}.

These successful checks raise a number of questions which are the motivations
of our investigation. The most compelling one is probably the following:  
how comes that
a saddle point approximation, in which only quadratic terms are retained
in the expansion of the action, is able to give the correct result?
In order to answer this question  and  extend 
previous computations,
we are led to consider the instanton calculus in
the framework of the topological twisted  version  
of $N=2$ SYM, 
the so--called Topological Yang--Mills theory (TYM) \cite{witten}.
As we will show later,  the ``traditional'' instanton calculus 
(the semiclassical expansion) and computations
carried on in the TYM framework give the same result. 
We will  extend the standard framework of TYM to encompass the
case in which the scalar field 
acquires a vacuum expectation value (v.e.v.).
To this end, 
we define a BRST operator for the ``extended'' TYM and find how it 
acts on the moduli space of 
(anti--)self--dual gauge connections, $\mathscr{M}^+$ ($\mathscr{M}^-$),
realized via the 
ADHM construction \cite{adhm}. 
From a geometrical point of view, this
BRST operator is on $\mathscr{M}^+$ the exterior 
derivative. We exploit this fact
to show that  all correlators made 
of insertions of gauge invariant polynomials in the scalar $\phi$, 
both with vanishing and non--vanishing  v.e.v.,  
can be written as total 
derivatives on  $\mathscr{M}^+$; 
an example is given by the modulus $u=< \Tr \phi^2 >$, which 
parametrizes the space of quantum vacua of the  $N=2$ SYM theory
with gauge group $SU(2)$.
In turn this result means that the only relevant
contributions to this class of correlators 
come from $\partial\mathscr{M}^+$.
To have an idea of what  the behavior of boundary terms is, we can look at
the compactified moduli space
$\overline\mathscr{M}_k$. In  \cite{dk} it was shown that  
$\partial{\overline\mathscr{M}}_k$ can be decomposed into 
the union of lower moduli spaces, so that one can write 
\beq
\overline\mathscr{M}_k=\mathscr{M}_k\cup\Rq\times\mathscr{M}_{k-1}\cup 
S^2\Rq\times\mathscr{M}_{k-2}\ldots\cup S^k\Rq
\ \ ,
\eeq
where $S^i\Rq$ denotes the $i^{th}$ symmetric product of points of $\Rq$.
The topological charge density, of winding number $k$, in 
$S^l\Rq\times\mathscr{M}_{k-l}$ is 
\beq 
|F_k|^2=|F_{k-l}|^2+\sum_{i=1}^l 8\pi^2\delta (x-y_i)
\label{occam3}
\ \ ,
\eeq
where $y_i\in S^i\Rq$ are the centers of the instanton. 
The Dirac delta functions 
are the contributions of zero size instantons, 
which ``factorize'' in
a fashion similar to a dilute gas approximation. 
This circumstance can lead to recursion relations of the type found  in  
\cite{marco,hoker}, thus  simplifying instanton calculations.

In this letter we briefly report on a series of calculations which establish
the formal equivalence between the coefficients 
of the Seiberg--Witten prepotential computed in the framework of 
the constrained instanton calculus and those computed 
in its TYM counterpart.
A longer report with all the details of our computations will be presented 
elsewhere \cite{bftt}.

\section{Topological Yang--Mills Theory and Instanton Moduli Spaces}
The relation between supersymmetric and topological theories
shows up when one observes that in the former there exists 
a class of position--independent Green's functions \cite{akmrv}. 
If one redefines
the generators of the 4--dimensional rotation group  in $\Rq$
in a suitably twisted fashion,
the $N=2$ SYM theory
gives rise to the TYM theory. 
With respect to this twisted Lorentz group,
SUSY charges decompose
as a scalar $Q$, a self--dual 
antisymmetric tensor $Q_{\mu\nu}$ and a vector
$Q_\mu$.
The field content  of $N=2$ SYM consists of 
a gauge field $A_\mu$, fermions  
$\lambda_{\alpha}^{\dot A}$ and
$\bar{\lambda}_{\dot\alpha}^{\dot A}$ 
($\dot{A}=1,2$ and   $\alpha , \dot{\alpha}=1,2$ are 
supersymmetry and spin indices respectively),
and a complex scalar field $\phi$, all in the adjoint of the gauge group
(that we take to be $SU(2)$).
Under the twist the fermionic degrees of freedom are reinterpreted 
in the following way:  
$\bar{\lambda}_{\dot\alpha}^{\dot A}\rightarrow \eta \oplus \chi_{\mu\nu}$,
$\lambda_{\alpha}^{\dot A} \rightarrow \psi_\mu$,  
where the anticommuting fields $\eta$, $\chi_{\mu \nu}$, $\psi_{\mu}$ are 
respectively a scalar, a self--dual antisymmetric tensor  and a vector.

The scalar supersymmetry of TYM 
plays a major r\^{o}le, in that it is 
still an invariance  of the theory  when this  is formulated 
on a generic (differentiable) four--manifold $M$. 
The Ward identities related to $Q$ implies that 
{\it all} the observables, including 
the partition function itself,  are 
topological invariants, in the sense that they are independent of 
the metric on $M$ \cite{witten}. Moreover, $Q$ 
has the crucial property of being nilpotent modulo gauge 
transformations; 
this allows one to interpret it as a BRST--like charge. 
In fact,  in order to have a strictly
nilpotent BRST charge, one needs to include gauge transformations
with the appropriate ghost $c$ \cite{bs}.
Defining $s=s_g + Q$, where
$s_g$ is the usual BRST operator associated to the gauge symmetry,  
the BRST transformations for the fields read 
\beqa
\label{BRST}
&&sA=\psi-Dc\ \ ,\nonumber\\
&&s\psi=-[c,\psi]-D\phi\ \ ,\nonumber\\
&&sc=-{1 \over 2} [c,c]+\phi\ \ ,\nonumber\\
&&s\phi=-[c,\phi]\ \ ,
\eeqa
whereas the anti--fields $\chi_{\m\n}$, $\bar\phi$ and $\bar c$ 
transform  under the BRST symmetry  as 
\beqa 
&& s\chi_{\m\n} = B_{\m\n}\ \ , \nonumber\\
&& s \bar\phi = \eta\ \ , \nonumber\\
&& s \bar c = b\ \ .
\label{anti-fi}
\eeqa 
$B_{\m\n}$, $\eta$ and $b$ are 
Lagrangian multipliers which transform as  
$s B_{\m\n} = 0$, $s\eta = 0 $, $s b = 0$. 
This algebra can be interpreted  as the definition and the Bianchi identities
for the curvature $\widehat{F}=F+\psi+\phi$ 
of the connection $\widehat{A}=A+c$ on 
the universal
bundle $P\times {\cal A}/{\cal G}$ ($P, {\cal A}, {\cal G}$ are 
respectively the principal bundle over $M$, the space of connections and
the group of gauge transformations).
The exterior derivative on the manifold $M\times {\cal A}/{\cal G}$
is given by $\widehat{d}=d+s$. 

It is important to distinguish two cases:
\begin{itemize}
\item[\bf 1.] trivial boundary conditions ($\phi=0$), and
\item[\bf 2.] non--trivial boundary conditions 
$\lim_{|x| \to \infty} \phi \equiv {\cal A}_{00}=v\sigma^{3}/2i$, $v\in \C$, 
\end{itemize}
for the field $\phi$  at $|x| \to \infty$.
In the first situation, the geometrical framework depicted above 
provides us with a
very nice explicit realization  of the BRST operator $s$ 
as the exterior derivative 
on the anti--instanton moduli space $\mathscr{M}^-$ \cite{witten2}.
This allows us to compute correlators of $s$--exact operators as integrals 
of forms on $\partial \mathscr{M}^-$ \cite{bftt}.
A problem arises when $\phi$ has 
non--trivial boundary conditions, 
since they are not compatible with the BRST algebra in 
(\ref{BRST}). This is because a non--vanishing  v.e.v. for $\phi$
implies the existence of a 
central charge $Z$ in the SUSY algebra 
which acts on the fields as a $U(1)$ transformation, so that 
$s=Q+s_g$ is not nilpotent; rather one gets
\beqa
&& (s_g + Q)^2 A = Z A\equiv - D\phi_Z \ \ , \nonumber\\
&& (s_g + Q)^2 \psi = Z \psi\equiv - [\phi_Z, \psi] \ \ , \nonumber\\
&& (s_g + Q)^2 \phi_Z = Z \phi_Z \equiv 0 \ \ .
\label{Z}
\eeqa
The scalar field $\phi_Z$ plays the r\^{o}le of
a gauge parameter and satisfies the equations
$D^2 \phi_Z = 0$,  $\lim_{|x| \rightarrow \infty}
\phi_Z  = {\cal A}_{00}$.
The $U(1)$ symmetry 
has to be included in the  appropriate extension of the BRST operator, 
through the introduction of a  
{\it global} ghost field $\L$ related to the
central charge symmetry.
We define an extended   BRST operator  as 
$s = s_g + Q - \lambda Z + {\partial\over{\partial\lambda}}$ \cite{hen},
where 
$\lambda$ is a fermionic parameter with ghost number $-1$ and canonical 
dimension zero,  such that $\lambda Z$ has the usual quantum numbers of a BRST 
operator.  One checks easily that $s$ is now nilpotent.
If we define  the ghost field 
$ \L \equiv \lambda \phi_Z$,
with the transformation 
$s \L = \phi_Z - [c,\L]$,
we can finally write the modified BRST algebra in the form 
\beqa
\label{brstZ}
&&sA=\psi-D(c+\L)\ \ ,\nonumber\\
&&s\psi=-[c+\L,\psi]-D\phi\ \ ,\nonumber\\
&&s(c+\L)=-{1\over 2}[c+\L,c+\L]+\phi \ \ ,\nonumber\\
&&s\phi=-[c+\L,\phi]
\ \ .
\eeqa
This algebra can be  derived  from (\ref{BRST})  by 
just making the replacement 
$c\longrightarrow c+\L$. Notice  that 
$\lim_{|x| \rightarrow \infty} (c + \L )= 
\lim_{|x| \rightarrow \infty} \L \equiv \lambda {\cal A}_{00}$.

A TYM action can be written as a pure 
gauge--fixing term \cite{bs} as follows, 
\beq 
S_{\rm TYM} = 2 \int d^4x~s \Tr \Psi \ \ ,
\label{s-tym}
\eeq
where the gauge--fixing fermion is chosen to be
$\Psi = \chi^{\m\n} F^+_{\m\n} - D^\m\bar\phi\psi_\mu + 
\bar c \partial^\m A_\m$. Explicitly
\beqa
S_{\rm TYM} &=& 
2\int d^4x \ \Tr \Bigl
( B^{\m\n}F_{\m\n}^+ - \chi^{\m\n}(D_{[\m}\psi_{\n]})^+ 
+ \eta D^\m \psi_\m + \nonumber\\
&& - \bar\phi ( D^2 \phi - [\psi^\m,\psi_\m] ) 
+ b \partial ^\m A_\m + \nonumber\\
&& + \chi^{\m\n} [c+\L,F_{\m\n}^+] - \bar\phi [c+\L,D^\mu\psi_\mu] - 
\bar c s(\partial^\m A_\m) \Bigr) + \nonumber\\
&& +
\int d^4x \ \partial^{\mu} s \Tr ( \bar\phi \psi_{\mu} )
\ \ ;
\label{ztym-act}
\eeqa
the last contribution   comes  from integrating by parts 
the term $\Tr (D^\m\bar\phi\psi_\m)$ in the gauge--fixing fermion.
The functional integration over anti--fields and 
Lagrangian multipliers 
projects onto the field subspace identified by the (zero--mode) equations
\beqa
&& F_{\m\n}^+ = 0 \ \ , \label{a-eq}\\
&& (D_{[\m}\psi_{\n]})^+ = 0 \ \ , 
\label{pippo}
\\
&& D^\m\psi_\m = 0 \ \ , 
\label{paperino}
\\
&& D^2\phi = [\psi^\m,\psi_\m] \ \ ,
\label{phieq}
\eeqa
where (\ref{phieq}) must be supplemented by appropriate 
boundary conditions on $\phi$.
Once the universal connection $\widehat{A}$ is given, the components 
$\left\{ F, \psi, \phi\right\}$ of
$\widehat{F}$ are in turn determined. $F$ is anti--self--dual,  and 
$\psi$ is an element of the 
tangent bundle $T_A\mathscr{M}^-$.

Notice that the solutions to
 (\ref{a-eq}), (\ref{pippo}), (\ref{paperino}) and (\ref{phieq})
supplemented by their boundary conditions
are  exactly the field configurations which are plugged into the 
functional integral in the context of the 
constrained instanton computational method \cite{aff} 
in  $N=2$ SYM \cite{fp,DKM,ft} as
{\it approximate} solutions of the $N=2$ equations of motion. 
As it is well--known, 
the action
obtained by twisting the $N=2$ SYM theory
({\it i.e.} the action of Witten's topological field theory \cite{witten})  
differs from (\ref{ztym-act})   
by some extra terms  which are BRST--exact
and correspond to 
a continuous deformation of the gauge--fixing \cite{bs}.
The Ward identities related to the BRST symmetry   
guarantee that  the two actions 
are  completely equivalent as for the computation of 
the observables of the theory: more precisely, 
{\it 
the Green's functions of  $s$--closed operators 
can be computed using  any  one of them obtaining the same result}.
This is why we used, with a slight abuse of language,
the same name TYM for both actions.

In the topological theory we are dealing with, 
functional integration reduces 
to an integration over  $\mathscr{M}^-$
and $T_A\mathscr{M}^-$. 
Both in the vanishing and in the non--vanishing v.e.v. case, 
the computation of the Green's functions we will consider
boils down to  integrating 
differential forms on the anti--instanton  moduli space, and 
their non--perturbative contribution will be symbolically 
represented by the formal expression
\beq
\label{prescription}
\left< fields \right> = \int_{\mathscr{M}^{-}}  \ 
\left[ (fields)\ e^{-S_{\rm TYM}} \right]_{zero-mode\  subspace}
\ \ .
\eeq
When $\phi$ has trivial boundary conditions,  $S_{\rm TYM}$ vanishes
on the zero--mode subspace.
For winding number $k=1$,  the top form on the 
(8--dimensional) anti--instanton moduli space is 
$\Tr \phi^2 (x_1)  \Tr \phi^2  (x_2)$, and one can compute  
the corresponding correlation function \cite{bftt}.
Since the BRST operator $s$ acts on $\mathscr{M}^-$
as the exterior derivative, we get the chain of equations
\cite{bftt}
\beq
<\Tr \phi^2 \Tr \phi^2 > = 
\int_{\mathscr{M}^-} \Tr \phi^2 \Tr \phi^2 = 
\int_{\partial \mathscr{M}^-} \Tr \phi^2 K_c = {1\over 2}
\ \ ,
\eeq
where the second equality follows from 
\beq
\Tr \phi^2 = s K_{c}\ \ , \ \ 
K_c = \Tr \left(csc + {2\over 3} ccc \right) \ \ ,
\eeq
an expression which parallels the well--known relation
\beq
\Tr F^2 = d K_{A}\ \ , \ \ 
K_A =
\Tr \left(AdA + {2\over 3} AAA\right) \ \ .
\eeq
The case in which $\phi$ has non--trivial boundary conditions
requires  a more detailed analysis.
The functional measure is in fact different from 1, 
since  now the  surface contribution 
\beq
S_{\rm inst} = 2\int d^4x \ \partial^{\mu} s \Tr ( \bar\phi \psi_{\mu} )
\ \ 
\label{DKM}
\eeq
is non--vanishing.
This gives rise to
non--trivial correlation functions which get contribution 
from topological sectors of {\it any} instanton number $k$;
the reason for that can be traced back to the fact that 
$\exp{(-S_{\rm inst} )}$ 
acts as a generating
functional for differential forms on $\mathscr{M}^-$.
The most interesting example  is 
$u(v) = \langle \Tr \phi^2 \rangle$, which  
parametrizes the quantum vacua of the $N=2$ theory, thus 
playing a prominent r\^{o}le in the context of the Seiberg--Witten 
solution for the $N=2$ Wilsonian action. 
In the next section we will sketch how the formal equation  
(\ref{prescription}) can be put to work 
in order to compute $u(v)$. 


\section{The ADHM Construction and Instanton Computations}
Before doing so, we briefly recall  some basic elements 
of the ADHM construction \cite{adhm}, which provides us with a 
parametrization of the  moduli space of 
self--dual connections (instantons)%
\footnote{In the previous section we have conformed to 
the standard convention in 
topological field theories of taking 
the gauge curvature to be anti--self--dual. Unfortunately
the literature on instanton calculus 
adopts the opposite convention (self--dual), to which 
we will conform from now on.
In the following we will 
use the definitions and conventions of Sec. II of \cite{ft}.},
$\mathscr{M}^+$. 
The dimension of $\mathscr{M}^+$ is $8k$; however the ADHM description 
is given in terms of a redundant set of 
parameters, and   its reparametrization symmetries
will play a major  r\^{o}le in the following. 
We will show that it is possible to realize 
the BRST algebra {\it directly} on the instanton moduli space, 
without having to solve any field equation.
This way the construction acquires a  geometrical meaning 
and stands on its own. 

The ADHM construction, 
which gives all   $SU(2)$  self--dual connections,
is purely algebraic and we find it more 
convenient to use quaternionic notations. The points, 
$x$, of the 
quaternionic space $\bb{H}\equiv \C^2\equiv\Rq$ 
can be conveniently
represented in the form $x=x^\mu \sigma_\mu$, with 
$\sigma_\mu=(i\sigma^b , \uno_{2\times 2}), b=1,2,3.$ 
The $\sigma^b$'s 
are the Pauli matrices, and
$\uno_{2\times 2}$ is the two--dimensional identity matrix.
To construct a self--dual connection 
of winding number $k$,  let us introduce a $(k+1)\times k$ 
quaternionic matrix linear in $x$, 
$\Delta=a+bx$, where  $a$ has the generic form 
\beq
\label{salute}
a=\pmatrix{w_1&\ldots&w_k\cr{}&{}&{}\cr{}&a'&{}\cr{}&{}&{}}\ \ ;
\eeq
$a'$ is a $k\times k$ quaternionic matrix, and
$b$ can be cast into the so--called ``canonical" form
\be
\label{canonical}
b=-\pmatrix {0_{1\times k}\cr\uno_{k\times k}} 
\ \ .
\ee
The (anti--hermitean) gauge connection can be written in the form
\be
A=U^\dagger d U
\ \ ,
\label{trecinque}
\ee
where $U$ is the solution to the equation 
$\Delta^\dagger U = 0$, with the  constraint
$U^\dagger U =\uno_{2\times 2}$ which  
 ensures that $A$ 
is an element  of the Lie algebra of the $SU(2)$ gauge group.
The condition of self--duality on the field strength of 
(\ref{trecinque}) is imposed by restricting the matrix $\Delta$ 
to obey
\be
\label{bos}
\Delta^\dagger\Delta=(\Delta^{\dagger}\Delta)^{T}
\ \ ,
\ee 
where the superscript $T$ stands for transposition of the
quaternionic
elements of the matrix (without transposing the quaternions 
themselves).
With the choice  (\ref{canonical}) for $b$, 
the instanton moduli space $\mathscr{M}^+$
is described in terms of a  
redundant set of  $8k + k(k-1)/2$ 
ADHM collective coordinates.
$A$ is invariant under the reparametrizations 
$\Delta\rightarrow  Q\Delta R$, 
where $R\in O(k)$, and  
\beq
Q = \pmatrix{q&0&\ldots&0\cr
0&{}&{}&{}\cr \vdots&{}&R^T&{}\cr
0&{}&{}&{}}
\ \ , 
\eeq
$q\in SU(2)$.
We are now ready to 
work out  the ADHM expression for $\widehat{A} = A + c + \L$ as an
educated extension of  (\ref{trecinque}).
We write 
\beq
\widehat{A} = U^\dagger (d+s+ {\cal C}) U \ \ ,
\label{apiuc-adhm}
\eeq
where ${\cal C}$ is the connection associated to the 
reparametrizations  of the ADHM construction, under which 
it undergoes the transformation 
${\cal C}\rightarrow  Q ({\cal C} + s ) Q^\dagger$.
Every expression should be covariant
with respect to this  reparametrization symmetry group;
this implies that ordinary derivatives on $\mathscr{M}^+$ 
have to be replaced by
covariant ones,  and $s$ by 
its covariant counterpart
$s + {\cal C}$.

We now construct the announced realization of the 
BRST algebra on  instanton moduli space. 
It will emerge as the most general 
set of deformations of the ADHM data $\Delta$ compatible
with the constraints (\ref{bos}).
As a bonus,  we will get a recipe 
to compute ${\cal  C}$. 
The central point is that gauging the reparametrization
symmetries  of the ADHM  description of  $\mathscr{M}^+$
is precisely what is required in order to make it possible 
to write BRST transformations of $\Delta$
consistent  with (\ref{bos}).
To show this, let us now start by performing an infinitesimal  
scalar variation (that we call $s$ for obvious reasons)
of (\ref{bos}). We get 
\beq
\label{sbos}
(s\Delta)^{\dagger}\Delta+\Delta^{\dagger}s\Delta=[(s\Delta)^{\dagger}
\Delta]^T+(\Delta^{\dagger}s\Delta)^T
\ \ , 
\eeq
which should be read as an equation for $s \Delta$.  
Its solution can be written as 
\beq
\label{sdelta}
s\Delta={\cal M}-{\cal C}\Delta\ \ ,
\eeq
where ${\cal M}$ is {\it defined} as the matrix which satisfies 
\beq
\label{fconstr}
\Delta^{\dagger}{\cal M}=(\Delta^{\dagger}{\cal M})^T \ \ .
\eeq
Notice that ${\cal M}$ is what in standard instanton calculations
would be called the fermionic  zero--mode  matrix.
We put ${\cal M}$ in a form which parallels the one for $a$ 
in (\ref{salute}),
\ie\
\beq
\label{salutem}
{\cal M}=
\pmatrix{\mu_1&\ldots&\mu_k\cr{}&{}&{}\cr{}&{\cal M}'&{}\cr{}&{}&{}}
\ \ , 
\eeq
${\cal M}'$ being a $k\times k$ symmetric quaternionic matrix.
The most general form for ${\cal C}$ consistent with all symmetries is 
\beq 
\label{parallel}
{\cal C}=\pmatrix{{\cal C}_{00}&0&\ldots&0\cr 0&{}&{}&{}\cr 
\vdots&{}&{\cal C}'&{}\cr
0&{}&{}&{}} \ \ ,
\eeq
where ${\cal C}'$ is a real antisymmetric $k\times k $ matrix;
${\cal C}_{00}$ is related to the asymptotic
behavior of the ghost $c+\L$ at infinity:
${\cal C}_{00} = \lambda  {\cal A}_{00}$.
If we now plug (\ref{sdelta}) into (\ref{sbos}), 
${\cal C}$ is determined by the equation 
\beq
\Delta^\dagger ({\cal C} + {\cal C}^\dagger) \Delta = 
\Big[ \Delta^\dagger ({\cal C} + {\cal C}^\dagger) \Delta \Bigr]^T
\ \ .
\eeq
${\cal C}^\prime$ is a 1--form 
expanded on a basis of differentials of the bosonic moduli. 
If one substitutes back the computed expression for ${\cal C}$ into 
the first of (\ref{sdelta}), then the ${\cal M}$'s
become in turn  1--forms 
on instanton moduli space $\mathscr{M}^+$.
As a specific example, let us consider the $k=2$ case in detail
(for $k=1$ the connection ${\cal C}^\prime$ vanishes).
The ADHM bosonic matrix $\Delta$ reads
\beq
\Delta=\pmatrix{w_1 & w_2\cr x_1 - x & a_1\cr a_1 & x_2 - x}
=\pmatrix{w_1 & w_2\cr a_3 & a_1\cr a_1 & -a_3}+b(x-x_0) \ \ ,
\label{f.14}
\eeq
where $x_0=(x_1+x_2)/2$, $a_3=(x_1-x_2)/2$. 
The solution to the bosonic constraint (\ref{bos}) is simply given by  
\beq 
\label{natale}
a_1=\frac{1}{4|a_3|^2}a_3(\bar{w}_2 w_1-\bar{w}_1w_2 +\Sigma)
\ \ ,
\eeq
where $\Sigma$ is an arbitrary real parameter related
to  the left--over $O(2)$ reparametrization invariance
(which can be exploited to put $\Sigma$ to zero).
Setting $H=w_1^2 + w_2^2 + 4(a_1^2 + a_3^2)$, one then finds
${\cal C}_{00} = \lambda  {\cal A}_{00}$, 
${\cal C}^\prime_{11} =
{\cal C}^\prime_{22} = 0$, and
\beq
{\cal C}^\prime_{12} =
-{\cal C}^\prime_{21}=\frac{1}{H}\bigg[
\bar{w}_1(s+{\cal C}_{00})w_2-
\bar{w}_2(s+{\cal C}_{00})w_1+ 2(\bar{a}_3 sa_1-\bar{a}_1 sa_3)\bigg]
\ \ .
\eeq

We now consider the fermionic constraint
(\ref{fconstr}).  Its $s$--variation  gives 
\beq
\label{sferm}
({\cal M}^{\dagger}+\Delta^{\dagger}{\cal C}){\cal M}-
[({\cal M}^{\dagger}+\Delta^{\dagger}{\cal C}){\cal M}]^T=
(\Delta^{\dagger}s{\cal M})^T-\Delta^{\dagger}s{\cal M}\ \ ,
\eeq
which should be thought of as an equation for 
$s{\cal M}$. 
Its  most general solution can be cast into the form 
$s{\cal M}={\cal A}\Delta-{\cal C}{\cal M}$,
where ${\cal A}$ has an expression which parallels (\ref{parallel}) 
and must satisfy the constraint 
\beq
\label{eqnA3}
\Delta^{\dagger}{\cal A}\Delta-(\Delta^{\dagger}{\cal
A}\Delta)^T=({\cal M}^{\dagger}{\cal M})^T-{\cal M}^{\dagger}{\cal
M}\ \ . 
\eeq
We want now to clarify the relation between 
${\cal A}$ and ${\cal C}$. 
To this end, let us perform one more $s$--variation of 
$s\Delta$ and  $s{\cal M}$; 
after a little algebra we get
\beqa
s^2 \Delta &=& \Bigl( {\cal A}- s{\cal C} - {\cal C} {\cal C}\Bigr)\Delta 
\ \ ,
\nonumber\\
s^2 {\cal M}&=&  \Bigl( {\cal A}- s{\cal C} - {\cal C} {\cal C}\Bigr) {\cal M}
+ \Bigl( s {\cal A} + [ {\cal C}, {\cal A}]\Bigr) \Delta 
\ \ .
\eeqa
The nilpotency of the BRST operator $s$ implies 
\beqa
\label{sera1}
&& {\cal A}- s{\cal C} - {\cal C}{\cal C} = 0
\ \ ,
\\
\label{sera2}
&&s {\cal A} + [ {\cal C}, {\cal A}] = 0
\ \ ;
\eeqa
therefore it becomes possible 
to consistently interpret (\ref{sera2}) as the Bianchi identity for 
${\cal A}$,  seen as the field--strength of ${\cal C}$ as per
(\ref{sera1}). 
Finally, we note that 
the variation of  (\ref{eqnA3}) simply gives an identity. 

Summarizing, we have found that consistency between  the 
$s$--variation  of the bosonic ADHM matrix $\Delta$ and the 
constraint (\ref{bos})  yields  
\beqa
\label{azzarolina}
s\Delta&=&{\cal M}-{\cal C} \Delta\ \ ,
\nonumber\\
s{\cal M}&=&{\cal A} \Delta-{\cal C}{\cal M}\ \ ,
\nonumber\\
s{\cal A}&=&-[{\cal C},{\cal A}]\ \ ,
\\
s{\cal C}&=&{\cal A}-{\cal C} {\cal C}\ \ ,
\nonumber
\eeqa
where $\Delta$ and ${\cal M}$ satisfy 
(\ref{bos}) and  (\ref{fconstr}) respectively. 
The set of equations (\ref{azzarolina}) is our main result as it yields  
an  explicit 
realization of the BRST algebra on the instanton moduli space. 

Two observations are in order. 
First, 
using (\ref{BRST}) or (\ref{brstZ}) 
we can work backward and deduce the explicit expressions for 
$\psi$, $\phi$, getting \cite{bftt} 
\beqa 
\label{brstfields}
\psi&=&U^{\dagger}{\cal M}f(d\Delta)^{\dagger}U+
U^{\dagger}(d\Delta)f{\cal M}^{\dagger}U\ \ ,
\nonumber\\
\phi&=&U^{\dagger}{\cal M}f{\cal M}^{\dagger}U+U^{\dagger}{\cal A}U
\ \ ,
\eeqa 
which coincide with the known solutions of the equations 
(\ref{pippo}), (\ref{paperino}), (\ref{phieq}) 
(in (\ref{brstfields}) we set 
$f\equiv(\Delta^\dagger \Delta)^{-1}$).
When inserted into (\ref{DKM}),
they yield exactly the supersymmetric multi--instanton action 
with non--zero v.e.v. for the scalar 
\cite{bftt} previously obtained in  \cite{DKM}.
Second, (\ref{sdelta}) gives us the opportunity  to discuss the
issue of the instanton measure in our framework. 
Let us call 
$\{ \widehat{\Delta}_{i}\} $ ($\{ \widehat{\cal M}_i\} $), $i=1,\ldots, p$
where $p=8k$, a basis of (ADHM) coordinates on 
$\mathscr{M}^+$ ($T_A\mathscr{M}^+$).
(\ref{sdelta}) then yields 
$\widehat{\cal M}_i = s \widehat{\Delta}_i + (\widehat{{\cal C} \Delta})_i$.
Therefore, 
the $\widehat{\cal M}_{i}$'s and the $s\widehat{\Delta}_{i}$'s are related 
by 
$\widehat{\cal M}_{i} = K_{ij} ( \widehat{\Delta} ) s\widehat{\Delta}_{j}$, 
where $K_{ij}$ is  a (moduli--dependent) linear transformation, 
which is completely known 
once the explicit expression for ${\cal C}$ is 
plugged into  $\widehat{\cal M}_{i}$.
After projection onto the zero--mode subspace,
{\it any polynomial in the fields becomes  a well--defined
differential form on} $\mathscr{M}^+$.
A generic  function on the zero--mode subspace
can be  written as  
\beqa
\label{inte1}
g( \widehat{\Delta}, \widehat{\cal M}) & = & 
g_{0} (\widehat{\Delta} )+ 
g_{i_1} (\widehat{\Delta} ) \widehat{\cal M}_{i_1} + 
{1\over 2!}
g_{i_1 i_2}(\widehat{\Delta}) \widehat{\cal M}_{i_1}\widehat{\cal M}_{i_2} +
\ldots 
\nonumber \\
&+&
{1\over p!}
g_{i_1 i_2 \ldots i_p} (\widehat{\Delta}) 
\widehat{\cal M}_{i_1}\widehat{\cal M}_{i_2}\cdots \widehat{\cal M}_{i_p}
\ \ ,
\eeqa
the coefficients of the expansion 
being totally antisymmetric in their indices.
It then follows that 
\be
\int_{\mathscr{M}^+}g( \widehat{\Delta}, \widehat{\cal M}) =
\int_{\mathscr{M}^+}  s^p  \widehat{\Delta} \  |{\rm det} K| 
g_{1 2 \ldots p} (\widehat{\Delta} )
\ \ ,
\eeq
where $s^p  \widehat{\Delta} \equiv \prod_{i=1}^{p} s \widehat{\Delta}_i$. 
This  formula is an operative tool to calculate physical amplitudes,
and the determinant of $K$ naturally stands out as 
the instanton integration measure for $N=2$ SYM theories. 
In standard instanton calculations 
this important ingredient  is obtained  as
a ratio of bosonic and fermionic zero--mode Jacobians.
In our framework it emerges instead in a geometrical and 
straightforward  way, 
{\it without the need of any computations of ratios of determinants nor 
of any knowledge of the explicit expressions of bosonic and fermionic 
zero--modes.}
The only ingredient is the connection 
${\cal C}$.
The instructive exercise of  
computing  $K$ and its determinant (\ie\ the instanton measure)
in the cases
of winding number equal to one and two \cite{bftt}
gives results which agree with previously known formulae
\cite{DKM}.

To make clear how the strategy for computing 
instanton--dominated correlators works
in our set--up, 
we now focus the attention on the Green's function 
$\left< \Tr \phi^2 \right>$, 
which is relevant for the computation of the 
Seiberg--Witten prepotential \cite{marco,ft}. 
In \cite{ft} it was shown that the 
topological sector of winding number $k$ contributes to it a 
\beq
\label{elena}
<{\Tr\phi^2\over 8\pi^2}>_{k} =
- k \int_{\mathscr{M}^+\backslash \{x_0\}}
e^{- [S_{\rm inst}]_{k}}
\ \ ,
\eeq
where $\mathscr{M}^+\backslash \{x_0\}$ is the
``reduced''  moduli space obtained 
after first integrating over the instanton center
$x_0$; the dimension of $\mathscr{M}^+\backslash \{x_0\}$  
is  $4n$, where $n=2k-1$. 
Let us call $\{\widetilde{\Delta}_i\}$, 
$i=1,\ldots,n$   a set of ADHM data 
for $\mathscr{M}^+\backslash \{x_0\}$,   
and $\widetilde{\cal M}_i$, $i=1,\ldots,n$
its fermionic counterpart. 
The exponential of the fermionic part $[S_F]_k$ 
of $[S_{\rm inst}]_k$ can now be expanded in powers. 
Under the integration over the reduced moduli space 
$\mathscr{M}^+\backslash \{x_0\}$, the only surviving 
term of the expansion will come from 
the top form on $\mathscr{M}^+\backslash \{x_0\}$.
It is crucial to remark that all the terms containing  ${\cal C}_{00}$
do not contribute to the amplitudes
since the parameter $\lambda$  does not 
belong to the moduli space. 
After some algebra one finds 
\beq
\label{tamara3}
<{\Tr\phi^2\over 8\pi^2}>_{k}
=
-k \cdot (32\pi^2)^{2n}  
\int_{\mathscr{M}^+\backslash \{x_0\}}s^{4n}\widetilde{\Delta}\ 
\det h  |\det K | e^{-\left[S_{B}\right]_{k}}
\ \ ,
\eeq
where $[S_B]_k$ is 
the bosonic part of the instanton action, and 
$h$ is defined through the formula 
$[S_F]_k=
8\pi^2 \overline{\widetilde{\cal{M}}}_{i}^{\dot{A}\alpha}
(h_{ij})_{\alpha}{}^{\beta}
(\widetilde{\cal M}_{j})_{\beta\dot{A}}$
($i,j=1,\ldots,n$).
Explicit computations of 
$ <\Tr\phi^2 /  8\pi^2>_{k}$  for  $k=1$ and $k=2$ 
from (\ref{tamara3}) 
give complete agreement with 
previously known results \cite{fp,DKM,ft}. 
Details will be presented elsewhere \cite{bftt}.

A new interesting possibility now arises  \cite{bftt}
observing that it is possible to write the action $S_{\rm inst}$  
as the $s$--variation of a certain function of the moduli,
more precisely in the form 
\beq
\label{act4exp}
\left[S_{\rm inst}\right]_{k}
=\left[ S_B+S_F\right]_{k}=4\pi^2 s\bigg\{\Tr\Big[\bar v(\sum_{i=1}^{k}\mu_i 
\bar w_i-w_i\bar{\mu}_i)\Big]\bigg\}
\ \ ,
\eeq
with $s [S_{\rm inst}]_k=0$.
As explained above, in order to compute (\ref{elena}), one has to 
extract   from $\exp{( - [ S_{\rm inst}]_k )}$ 
the top form on $\mathscr{M}^+\backslash \{x_0\}$, which  reads 
\beq
\label{act4exp2}
\left. e^{-[S_{\rm inst}]_k}\right|_{top\ form}=
4\pi^2 s\left\{\Tr\Big[\bar v(\sum_{i=1}^{k}\mu_i \bar w_i-
w_i\bar{\mu}_i)\Big] ([S_F]_k)^{4k-3}([S_B]_k)^{-4k+2}
\bigg(1-e^{-[S_B]_k}\sum_{l=0}^{4k-3}{([S_B]_k)^l\over l!}\bigg)\right\}
\  .
\eeq
In fact, (\ref{act4exp2}) enables one to write  $<\Tr\phi^2>_{k}$
as an integral over the boundary of the  instanton moduli space. 
The circumstance that Green's functions  can   be  
computed in principle on the boundary of $\mathscr{M}^+$ may greatly help in 
calculations,  since instantons on $\partial\mathscr{M}^+$ obey a kind
of dilute gas approximation \cite{bftt}, as emphasized in the Introduction. 
We leave to future work the computations with $k>1$, 
limiting here our attention  to the $k=1$ case.

From the analyses of \cite{dk,Freed}, 
it is known that the boundary of the $k=1$ moduli  space
consists of instantons of zero ``conformal'' size;
this means that if we projectively map the Euclidean flat space 
$\Rq$ onto a 4--sphere $S^4$, the boundary of the
corresponding transformed $k=1$ instanton moduli space 
is given by instantons of zero
conformal size $\tau$, where $\tau$ we is
obtained from $|w|$ through a projective
transformation ($|w|$ itself does not yield  a globally defined
coordinate on the $S^4$ instanton moduli space). 
In terms of the size $|w|$ of the $\Rq$ instanton, the 
$\tau\to 0$ limit corresponds to $|w|\rightarrow 0,\infty$. 
Specializing (\ref{act4exp2}) to  $k=1$ and inserting it in 
(\ref{elena})  we get
\beqa
\label{k=1bordo}
<{\Tr\phi^2\over 8\pi^2}>_{k=1}
&=& - 4\pi^2 \int_{\mathscr{M}^+\backslash \{x_0\}}
s \biggl\{ 
\Tr\Big[\bar v(\mu \bar w- w\bar\mu)\Big] [S_F]_{k=1} {1\over [S_B]_{k=1}^2}
\cdot
\nonumber \\
&&\cdot
\Bigl( 1- e^{-[S_B]_{k=1}} - [S_B]_{k=1} e^{-[S_B]_{k=1}} \Bigr)\biggr\}
\nonumber \\
&=&
- \left.{8\pi^2\over v^2}
\Big(1-e^{-4\pi^2|v|^2|w|^2}-4\pi^2 |v|^2|w|^2
e^{-4\pi^2 |v|^2|w|^2}\Big)\right|_{|w|=0}^{|w|=\infty}
\nonumber \\
&=&-{8\pi^2\over v^2}
\ \ .
\eeqa
In the second equality of (\ref{k=1bordo}) 
use of the Stokes' theorem has been made 
to express $\left<\Tr\phi^2/  (8\pi^2)\right>_{k=1}$ as an integral over
the boundary  of 
$  \mathscr{M}^+\backslash \{x_0\} $,  which for
$k=1$ is $\partial \R^{+}\times S^3/\Z_2$. 
Again, (\ref{k=1bordo}) gives the correct answer.

We believe that this approach 
provides a natural and simplifying  framework for studying 
non--perturbative effects in supersymmetric  gauge theories.

\section*{Acknowledgements}
We are particularly indebted to G.C. Rossi for many valuable
discussions over a long time, 
and for comments and suggestions 
on a preliminary version of this letter.
We are also grateful to 
D. Anselmi, C.M. Becchi, P. Di Vecchia, S. Giusto,
C. Imbimbo,  V.V. Khoze, M. Matone, 
S.P. Sorella and R. Stora for many stimulating conversations.
D.B. was partly supported by the Angelo Della Riccia Foundation.

\end{document}